\documentclass[vecphys]{svmult}

\usepackage{makeidx}         
\usepackage{graphicx}        
\usepackage{multicol}        
\usepackage[bottom]{footmisc}
\usepackage{cite}
\usepackage{url}

\makeindex             

\begin{document}

\title*{Tight-binding modeling of charge migration in DNA devices}
\titlerunning{Tight-binding models of charge migration in DNA}
\author{G.\ Cuniberti\inst{1}\and
E.\ Maci\'{a}\inst{2} \and A. Rodriguez\inst{3} \and R.A.\ R\"{o}mer\inst{4}}
\institute{Institute for Theoretical Physics, University of
Regensburg,
  D-93040 Regensburg, Germany \texttt{g.cuniberti@physik.uni-R.de} \and Departamento de
  Fisica de Materiales, Universidad Complutense de
  Madrid, E-28040 Madrid, Spain \texttt{emaciaba@fis.ucm.es} \and Dpto.\
  Matem\'atica Aplicada y Estad\'{\i}stica, EUIT. Aeron\'{a}utics U.P.M., PZA
  Cardenal Cisneros s/n, Madrid 28040, Spain
  \texttt{antonio.rodriguezm@upm.es}\and Department of Physics and
  Centre for Scientific Computing, University of Warwick, Coventry CV4
  7AL, United Kingdom \texttt{Rudolf.Roemer@warwick.ac.uk}}
%
%
\maketitle
\setcounter{minitocdepth}{3} \dominitoc
\begin{abstract}
Long range charge transfer experiments in DNA oligomers and the
subsequently measured --- and very diverse --- transport response of
DNA wires in solid state experiments exemplifies the need for a
thorough theoretical understanding of charge migration in DNA-based
natural and artificial materials. Here we present a review of
tight-binding models for DNA conduction which have the intrinsic
merit of containing more structural information than plain
rate-equation models while still retaining sufficient detail of the
electronic properties. This allows for simulations of transport
properties to be more manageable with respect to density functional
theory methods or correlated first principle algorithms.
\end{abstract}

\section{Introduction and motivation}
\label{sec-introduction}

Within the class of biopolymers, DNA is expected to play an
outstanding role in molecular electronics \cite{StaLT06}. This is
mainly due to its unique self-assembling \index{self-assembling} and
self-recognition \index{self-recognition} properties which are
essential for its performance as carrier of the genetic code. It is
the long-standing hope of many scientists that these properties
might be further exploited in the design of electronic circuits
\cite{EleS62,KelB99,KerBBS03,MerKPE99,ReiLS01}.
In the last decade of the 20th century, {\em transfer} experiments in
natural DNA in solution showed unexpected high charge transfer rates
\cite{MurAJG93,PriRB96,MegBG98,KelB99,TreHB01}. This suggested
that DNA might support charge transport. In contradistinction, {\em
  electrical transport} experiments carried out on single DNA molecules
displayed a variety of possible behaviors: insulating
\cite{BraESB98,StoVDD01}, semiconducting \cite{PorBVD00,CohNNP05}
and ohmic-like \cite{FinS99,YooHLP01,XuZLT04,KasKGRV01}. This
variation might be traced to the high sensitivity of charge
propagation in DNA to extrinsic (interaction with hard
substrates\index{hard substrates}, metal-molecule
contacts\index{metal-molecule contacts}, aqueous
environment\index{aqueous environment}) as well as intrinsic
(dynamical structure fluctuations\index{dynamical structure
fluctuations}, base-pair sequence\index{base-pair sequence})
influences. Recently, experiments on single poly(GC) oligomers in
aqueous solution~\cite{XuZLT04} as well as on single suspended DNA
with a more complex base sequence \cite{CohNNP05} have shown
unexpectedly high currents of the order of $100$--$200$ nA. Again
these results, if further confirmed, suggest that DNA molecules may
support rather high electrical currents given the right
environmental condition.

The theoretical interpretation of these recent experiments and, in a
more general context, the elucidation of possible mechanisms for
charge transport in DNA have not, however, been unequivocally
successful so far.
While {\em ab initio} calculations
\cite{ArtMSO03,CalDMG02,DifCZ04,GerCP02,BarCJL01,HubECS05,Sta04,Sta05,AdeWA03,MehA05}
can give at least in principle a detailed account of the electronic and
structural properties of DNA, the huge complexity of the molecule and
the diversity of interactions present preclude a complete treatment for
realistic molecule lengths. When interactions with counter ions and
hydration shells or vibrational degrees of freedom are to be
considered, the situation easily becomes untractable.
On the other hand, model-based Hamiltonian approaches to DNA
\cite{CunCPD02,JorBLM98,JorB02,RocBMK03,Roc03,UngS03,PalAHR04,ApaC05,ApaC05a,GutMC05a,GutMC05,KloRT05,Yam04,DorB03,MacR06,Mac06}
have been already been discussed in great detail and can play a
complementary role by addressing single factors that influence charge
transport in DNA. However, here it is of course clear neither  a-priori nor
a-posteriori (given the aforementioned experimental situation)
which model should be used. Somewhat mirroring the experimental
situation, a large variety of models exist and the results are not
necessarily consistent across different models.

In this contribution, we review tight-binding models of DNA which have
been proposed in the literature and argued to reproduce experimental
\cite{CunCPD02} as well as {\em ab-initio} results \cite{DavI04}.
We first concentrate on simple one- and two-channel models of DNA in
which the main transport mechanism is concentrated in the effects of
$\pi$-overlap in the base or base-pair sequences. The models are usually
constructed to take into account the HOMO-LUMO gap of the single base pairs
similar to many of the
DFT-based studies.
A main feature of the next class of models is the presence of sites
which represent the sugar-phosphate backbone\index{sugar-phosphate
backbone} of DNA but along which no electron transport is
permissible. These models construct a gap due to transversal
perturbation of the $\pi$-stack, i.e.\ even when the onsite energies
are constant.
The aim of this review is thus to explain the present state of affairs
in the tight-binding-model-based approach and we will be very brief on
many others aspects of the charge migration problem, as these are
already well treated in the other chapters of this book.

\section{The electronic structure of DNA}
\label{sec-electronicstructure}

\begin{figure}
  \centering
  \includegraphics[width=0.9\textwidth]{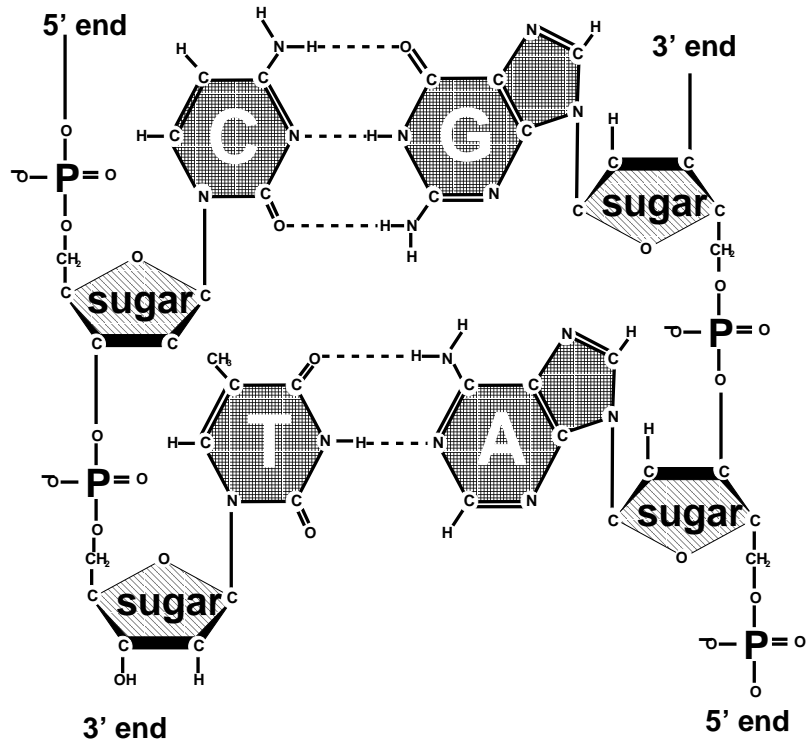}
  \caption{\label{fig-DNA}
    The chemical composition of DNA with the four bases Adenine (A),
    Thymine (T), Cytosine (C), Guanine (G) and the backbone. The
    backbone is made of phosphorylated sugars shown in light grey, the
    nucleobases are indicated in dark grey.}
\end{figure}

\begin{figure}
  \centering
  \includegraphics[width=0.7\textwidth,angle=-90,clip]{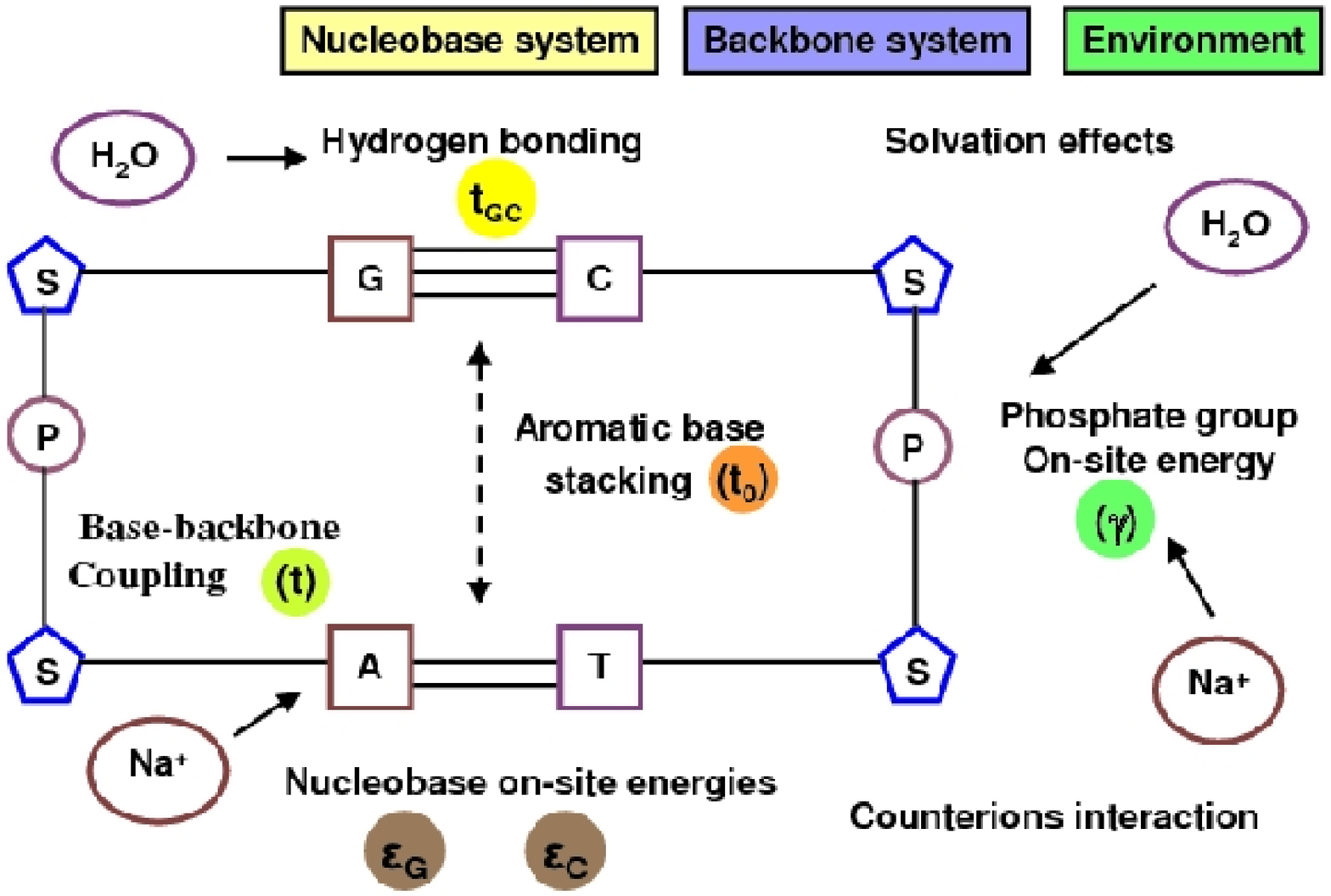}
  \caption{\label{energetics}
    Sketch illustrating the overall energetics of a double-stranded DNA
    chain.}
\end{figure}

DNA is a macro-molecule consisting of repeated stacks of bases formed by
either AT (TA) or GC (CG) pairs coupled via hydrogen bonds and held in
the double-helix structure by a sugar-phosphate backbone. In Fig.\
\ref{fig-DNA}, we show a schematic drawing.
The electronic energetics of a double-stranded DNA chain should take
into account three different contributions coming from (i) the
nucleobase system, (ii) the backbone system and (iii) the environment,
as it is sketched in Fig.\ \ref{energetics}.
Attending to the energies involved in the different interactions,
the resulting energy network\index{energy network} can be
hierarchically arranged, starting from high energy values related to
the on-site energies of the bases and sugar-phosphate groups
($8-12$~eV)\cite{YanZ02,Igu04} passing through intermediate energy
values related to the hydrogen bonding between Watson-Crick pairs
($\sim 0.5$~eV) \cite{YanZ02} and the coupling between the bases and
the sugar moiety ($\sim 1$~eV) \cite{Igu04} and ending up with the
aromatic base stacking low energies ($0.01-0.4$~eV)
\cite{YanZ02,VoiJBR01}. The energy scale of environmental
effects\index{energy scale of environmental effects} ($1-5$~eV) is
related to the presence of counter ions and water molecules,
interacting with the nucleobases and the backbone by means of
hydration, solvation and charge transfer processes. It is about one
order of magnitude larger than the coupling between the
complementary bases, and about two orders of magnitude larger than
the base stacking energies.

We emphasize that in many of the models to be reviewed later in this
chapter, simplified assumptions about these energy scales are
employed. Mostly, however, the ionization energies\index{ionization
energies} $\epsilon_{\rm G}=7.75$~eV, $\epsilon_{\rm C}=8.87$~eV,
$\epsilon_{\rm A}=8.24$~eV and $\epsilon_{\rm
  T}=9.14$~eV, \cite{SugS96,VoiJBR01,ZhaLHY02,YanWVW04,CauDL06} are taken
as suitable approximations to the onsite energetics at each base.

\section{Numerical techniques for charge transport in the quantum regime}
\label{sec-quantumtransport}

Before we in the following turn our attention to the variety of
simplified models which have been proposed to capture the essential
charge transport features of DNA, let us briefly recall some of the
techniques used to investigate these.

There are several approaches suitable for studying the transport
properties of quasi-one-dimensional tight-binding
models\index{tight-binding models} for long DNA and these can be
found in the literature on transport in solid state systems, or,
perhaps more appropriately, quantum wires\index{quantum wires}
\cite{HofS02}. Since the variation in the sequence of base pairs
precludes a general solution, one normally uses methods well-known
from the theory of disordered systems \cite{KraM93,RomS03}.  The
main advantage of these methods is that they work reliably (i) for
the relatively short DNA strands ranging from $13$ base pairs (as in
DFT studies \cite{PabMCH00}) up to 30 base pairs length which are
being used in the nanoscopic transport measurements \cite{PorBVD00}
as well as (ii) for somewhat longer DNA sequences as modeled in the
electron transfer results and (iii) even for complete DNA sequences
which contain, e.g.\ for human chromosomes up to 245 million base
pairs \cite{AlbBLR94}.
We measure the effectiveness of the electronic transport by various
measures such as the {\em localisation length}
$\xi$\index{localisation length}, participation
numbers\index{participation number}, etc. These roughly speaking
parameterise whether an electron is confined to a certain region of
the DNA (resulting in insulating behavior) or can proceed across the
full length $L$ of the DNA molecule (metallic behavior).

\subsection{Recursive Green function technique}
\label{sec-rgfm}

The first method one can use is the recursive Green function
approach\index{recursive Green function approach} pioneered by
MacKinnon \cite{Mac80,Mac85}. It can be used to calculate the dc and
ac conductivity tensors and the density of states (DOS) of a
$d$-dimensional disordered system and has been adopted to calculate
all kinetic linear-transport coefficients such as thermoelectric
power, thermal conductivity, Peltier coefficient and Lorenz number
\cite{RomMV03,CroRS06}. Briefly, the approach utilizes the advanced
and retarded Green's functions, ${\cal G}^{-}(E-i0^+)$ and ${\cal
  G}^{+}(E+i0^+)$, respectively, and the usual definition $\left[ (E\pm i
  0^+) \delta_{ij} - H_{ij} \right] G_{ij}^{\pm} = \delta_{ij}$, where
$G_{ij}^{\pm}$ is the matrix element $\langle i | {\cal G}^{\pm} | j
\rangle$ and $H_{ij}$ is similarly the matrix element of the
Hamiltonian \cite{CroRS06}. $\delta_{ij}$ denotes the Kronecker
$\delta$ between basis states $\{| i \rangle\}$. If contains only
nearest-neighbor connections, then these expressions can be written
interactively as
\begin{equation}
  \label{eq:rgfm}
  - H_{ii+1}G^{\pm}_{i+1j}= \delta_{ij} - \left[ (E\pm i
  0) \delta_{ij} - H_{ij} \right] G_{ij}^{\pm} + H_{ii-1}G^{\pm}_{i-1j}.
\end{equation}
Here $H_{ii\pm 1}$ are the terms in the Hamiltonian connecting slice
$i$ with its neighboring slices $i\pm 1$.  If we now reinterpret the
left index $i$ as a pseudo time, then we see that the future Green
function slice $i+1$ can be constructed by the present slice at $i$
and the previous slice at $i-1$. The method is well suited to study
coherent transport properties\index{coherent transport properties}
and can be extended to include incoherent processes\index{incoherent
transport properties} as well \cite{DamP90}.

\subsection{Transfer and transmission matrix approach}
\label{sec-tmm}

The next method of choice is the iterative transfer-matrix method
(TMM)
\cite{PicS81a,PicS81b,MacK83,KraM93,Mac94}\index{transfer-matrix
method} which allows us in principle to determine the localisation
length $\xi$ of electronic states in systems with varying cross
section $M$ and length $L > M$. This localization length describes
the decay of the wave function for transport along a quasi
one-dimensional system and $\xi$ may be used as a rough guide of the
extend of electronic states\index{extend of electronic states
}.

For disordered systems, typically a few million sites $L \gg M$ are
needed to achieve reasonable accuracy for $\xi$ \cite{KraM93}. However,
in the present situation we are interested in finding $\xi$ also for
DNA strands of typically only a few hundred or a few ten thousand
base-pair long sequences.  Thus in order to restore the required
precision, one modifies the conventional TMM and now performs the TMM
on a system of fixed length $L_0$. This modification has been
previously used \cite{FraMPW95,RomS97b,NdaRS04} and may be summarized
as follows:
After the usual forward calculation with a global transfer
matrix\index{global transfer matrix} ${\cal
  T}_{L_0}$, we add a backward calculation with transfer matrix ${\cal
  T}^{\rm b}_{L_0}$. This forward-backward-multiplication procedure is
repeated $K$ times. The effective total number of TMM multiplications
is $L_{\rm }=2KL_0$ and the global transfer-matrix is ${\tau}_{L_{\rm
}} = \left( {\cal T}^{\rm b}_{L_0} {\cal T}_{L_0}\right)^K$. It can be
diagonals as for the standard TMM with $K\rightarrow \infty$ to give
${\tau}^{\dagger}_{L_{\rm }} {{\tau}_{L_{\rm }}} \rightarrow \exp[ {\rm
  diag}(4KL_0/\xi_i)]$. The largest
$\xi_i$ for all $i=1, \ldots, M$ then corresponds to the localisation
lengths of the electron on the DNA strand and will be measured in units
of the DNA base-pair spacing ($0.34$ nm).
Let us emphasize that the above approach converges even for $L < \xi$.
However, in that case, values of $\xi$ clearly are dominated by
finite-size and boundary effects and their significance is no longer
quantitative, but qualitatively indicates extended states smeared out
over the finite system length $L$.
Last, the transmission coefficient $T_L(E)$, related to the Landauer
conductance\index{Landauer conductance} $g$ via $g=(2 e^2/h)
T_L(E)/(1-T_L(E))$ \cite{ButIL83,ButILP85,BagO89}, is defined in
terms of the matrix elements of $\tau_L$.

\subsection{Attaching leads}

Let us assume that, as a first approximation, we can consider a DNA
model in terms of a linear chain with an orbital per site, where
each lattice site represents a base pair. The ends of the chain are
connected to leads modeled as semi-infinite one-dimensional chains
of atoms with one orbital per site. Broadly speaking, one expects
the binding to metallic leads\index{metallic leads} would affect the
electronic structure of the molecule. If so, we should consider the
states belonging to the coupled molecular-metallic system rather
than those of the molecular subsystem alone \cite{EmbK98}. Thus we
shall consider henceforth that the coupling between the contacts and
the molecule is weak enough, so that the lead-molecule-lead
junction\index{lead-molecule-lead junction} can be properly
described in terms of three non-interacting subsystems
\cite{EmbK99,Kos02}, according to
\begin{eqnarray}
\label{eq:leads}
H &= &H_{\rm DNA} - t_{\rm Contact} \left( |0\rangle \langle 1| + |N\rangle \langle N+1| + h.c.\right)
\nonumber \\
& & \mbox{ }
+\sum_{k=0}^{-\infty } \varepsilon _{\rm Lead} |k\rangle\langle k| - t_{\rm Lead} |k-1\rangle\langle k| + h.c.
\nonumber \\
& & \mbox{ }
+\sum_{k=N+1}^{+\infty } \varepsilon _{\rm Lead} |l\rangle\langle l| - t_{\rm Lead} |l\rangle\langle l+1| + h.c. .
\end{eqnarray}
In Eq.(\ref{eq:leads}), $H_{\rm DNA}$ is the DNA
Hamiltonian\index{DNA Hamiltonian}, the second term describes the
DNA-lead contact, and the last two terms describe the contacts at
both sides of the DNA chain, where $N$ is the number of base pairs,
$\varepsilon _{n}$ are the on-site energies of the base pairs,
$t_{\rm Contact}$ is the hopping strength between the leads and the
end nucleotides, $\varepsilon _{\rm Lead}$ is the leads on-site
energy and $t_{\rm Lead}$ is their hopping term.

The Green function methods\index{Green function method} are
well-suited to include contact effects since their boundary
conditions at the contacts require specification of a suitable Green
function in the leads which can be chosen to model the geometry of
contacts. The TMM usually starts assuming a particle-like injection
of carriers into the transport channels and a proper treatments of
leads is lacking, but the extracted localization lengths at least
for long chains are largely independent of the exact choice.
Irrespective of numerical methods used, most earlier tight-binding
studies assumed perfect coupling to metallic leads or simply ignore
the issue altogether. The role of contact effects within the TMM
framework was recently reported for a poly(GACT) tetra-nucleotide in
Ref.\ \cite{MacTR05} in terms of two contact matrices which
explicitly take into account the presence of the $t_{\rm Contact}$
hopping integral. Depending on the value adopted for $t_{\rm
Contact}$, the obtained transmission coefficient does not reach in
general the full transmission condition $T_L(E)=1$ due to the
symmetry breaking related to the coupling of the G (T) end
nucleotides at the left (right) leads, respectively. This extreme
sensitivity is due to interference effects between the DNA energy
levels and the electronic structure of the leads at the metal-DNA
interface, and indicates that the {\em optimal} system configuration
for efficient charge transfer is determined by the resonance
condition  $t_{\rm Contact} =\sqrt{t\cdot t_{\rm Lead}}$. Quite
interestingly, one realizes that, due to resonance effects, a
stronger coupling to the leads not always result in a larger
conductance through the system, in agreement with the results
obtained by Guo and co-workers for the transmission coefficient of
poly(G)-poly(C) molecules making use of the Green function technique
\cite{ZhuKG04}. Subsequent works have exploited the existence of
this optimal charge injection condition to study the charge
migration efficiency through more realistic duplex chains (see the
contribution by Apalkov, Wang, and Chakraborty in this volume)

In general, modeling the geometry and bonding character of the
contact at the interface is a very delicate issue, since detailed
information on both the metal geometry and DNA chemical bonding at
the contacts is poorly known to date. Consequently, in most modeling
of the DNA-contact interface, the parameter $t_{\rm Contact}$ deals
with the tunneling probability between the frontier orbitals, thus
roughly encompassing bonding effects at the interface.
Recent transport experiments have shown that deliberate {\em chemical
  bonding} between DNA and electrodes is a prerequisite for achieving
reproducible conductivity results \cite{HarMAC003,ZhaAKC02,StoVDD01}.
Accordingly, the study of contact effects on the charge migration
efficiency is an important issue to be considered in realistic
models of DNA transport.

\section{Tight-binding model approaches}
\label{sec-modelapproaches}

The ab initio methods are clearly very powerful. However, from a
physics perspective, the question immediately arises if an even
simpler, effective model approach, might capture the essentials of
charge migration equally well. This strategy is known as the
tight-binding approach to DNA --- note that in this language the
term {\em tight-binding} is used somewhat differently from the
terminology of theoretical chemistry. This simple approach has been
used right from the start of the physics involvement in DNA
research. The idea is to capture the main path-ways of charge
migration along the DNA molecule stack in a simple model of {\em
sites} and {\em hopping strengths}. Charge transport along this
model is then described by simple tight-binding
orbitals\index{tight-binding orbital} on the sites and suitably
parametrized hopping onto neighboring sites. The advantage is this
approach should be clear: once the appropriate onsite
energies\index{onsite energy} and hopping strengths\index{hopping
strength} are known, much larger system sizes can be studied than
with the ab initio methods. The downside of course is that the
determination of the effective parameters\index{effective
parameters} and in particular the choice which to leave out
completely will be at least to some degree a matter of personal
preference and thus open to criticism.

\subsection{Importance of the DNA sequence: one-dimensional models}
\label{sec-wire}

\begin{figure}
  \centering
  \includegraphics[width=0.9\textwidth]{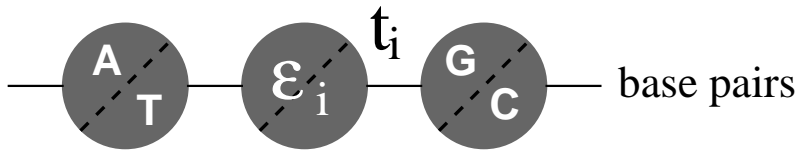}
  \caption{\label{fig-line} The line model for electronic transport
    along DNA corresponding to the Hamiltonian given in Eq.\
    (\ref{eq:wire-ham}). Lines denote hopping amplitudes and circles
    give the central (grey) nucleobase pairs.}
\end{figure}

The simplest TB model of the DNA stack can be constructed as a
one-dimensional model as given in Fig.\ \ref{fig-line}.
There is a single central conduction channel\index{central
conduction channel} in which individual sites represent a base-pair.
Every link between sites implies the presence of a hopping
amplitude. The Hamiltonian for this {\em wire} model $(H_{\rm
  W})$ is given by
\begin{eqnarray}
H_{\rm W} &=& \sum_{i=1}^{L}
     -  t_{i} |i \rangle \langle i+1| - t_{i-1} |i \rangle \langle i-1|
    + \varepsilon_i |i \rangle \langle i| \label{eq:wire-ham}
\end{eqnarray}
where $t_{i}$ is the hopping between nearest-neighbour sites $i,i+1$
along the central branch and we denote the onsite energy at each site
along the central branch by $\varepsilon_i$.  $L$ is the number of
sites/bases in the sequence.
For constant $t_i=t$, $\epsilon_i=0$ and open boundary conditions, the
spectrum of the model is given by
\begin{equation}
  \label{eq:wire-spectrum}
  E=-2 t \cos\left(\frac{\pi k}{L+1}\right)
\end{equation}
with $k=1, 2, \ldots L$.
For random choice of onsite energies or hopping strengths, this
model is well-known as the Anderson model\index{Anderson model}
\cite{And58} with diagonal or off-diagonal disorder and its
transport properties are governed by one-dimensional Anderson
localization\index{Anderson localization} \cite{BraK03}.

In order to use this Hamiltonian to model DNA, one needs to know the
appropriate parameters for onsite energies and hopping strengths
\cite{SugS96,VoiJBR01,ZhaLHY02,YanWVW04,CauDL06}, or, alternatively,
argues that mostly the statistical properties of these quantities
determine the transport.
For natural DNA sequences, a useful choice for the onsite energies
might be the ionization potentials mentioned in section
\ref{sec-electronicstructure}.
But since base-pairs are modelled by a single site, the DNA is
effectively described as a sequence of GC (identical to CG) and AT (or
TA) pairs with links between like (GC-GC or AT-AT) or unlike (GC-AT,
AT-GC) pairs. Thus the model parameters for the pairs should be
computed as suitable estimates based on the ionization potentials of
individual bases \cite{SugS96,VoiJBR01,ZhaLHY02,CauDL06,YanWVW04}.

Already such a simple model as (\ref{eq:wire-ham}) allows to study various
aspects of charge transport in DNA.
Electrical transport through invididual DNA molecules was studied in
Ref.\ \cite{LiY01}, using poly(G)-poly(C) DNA. Individual molecules
are coupled to external baths \cite{Datta99}, thus leading to
partial decoherence. Good agreement with the experimental results of
Ref.\ \cite{PorBVD00} is demonstrated.
A twist angle in the hopping parts of (\ref{eq:wire-ham}) is used in
Ref.\ \cite{YuS01} to model the effect of thermal fluctuations on
transport in DNA. The participation ratio\index{participation ratio}
is used to estimate the extent of the electronic states. Assuming
that inelastic effects due to temperature can be ignored, the paper
then computes the temperature dependence of the conductivity.
The transmission spectrum for a chain of poly(G)-poly(C) DNA molecules
was studied in Ref.\ \cite{ZhuKG04} where also disorder and contact
effects were taken into account. The model contains various parameters
according to the HOMO/LUMO structure of DNA. Furthermore, charging
effects, i.e.\ Coulomb blockade, were studied within a mean-field
approach.
For a DNA chain consisting of AT and GC pairs, Ref.\ \cite{ZhaU04}
investigates structural and dynamical disorder\index{dynamical
disorder}. Here, in addition to the onsite energies in
(\ref{eq:wire-ham}), also the hopping elements $t_i$ are chosen
according to the specific DNA sequence, which itself, however,
consists of random sequences of A,T,G,C nucleotides. It is shown
that both types of disorder can significantly influence the
transport properties.
Ref.\ \cite{Asa03} studies both (quasi-)coherent\index{coherent
transport} and incoherent transport\index{incoherent transport}
regimes using Landauer\index{Landauer formalism} and Kubo\index{Kubo
formalism} formalism via a continued fraction
approach\index{continued fraction approach} for poly(G)-poly(C) and
also poly(A)-poly(T) DNA chains.
Superexchange-like\index{superexchange} exponential length
dependence is found for the coherent and Ohmic-like behavior for the
incoherent regimes.

The next group of studies focuses on the influence that possible
correlations\index{correlations} in both artificial and natural DNA
sequences might have on the transport. Natural
$\lambda$-phage\index{$\lambda$-phage} DNA has been investigated in
Ref.\ \cite{Roc03} within a transfer-matrix
approach\index{transfer-matrix approach}. Transmission spectra are
shown to be very different from poly(G)-poly(C) DNA. The results are
argued to be roughly consistent with those from electron {\em
transfer} studies.
The influence of long-range correlations\index{long-range
correlations} in DNA sequences is studied in Ref.\ \cite{RocBMK03}.
Natural DNA of the first completely sequenced human chromosome 22
(Ch22) \index{human chromosome 22} is compared to articial sequences
such as random and Fibonacci sequences\index{Fibonacci sequence}. Is
is found that long-range correlations \index{long-range
correlations} induce coherent charge transfer over longer lengths
scales, at least for Ch22.
Ref.\ \cite{AlbVLD05} uses the same numerical method as Ref.\
\cite{RocBMK03} and corroborates the results for Ch22 by comparing
to a Rudin-Shapiro sequence\index{Rudin-Shapiro sequence}.
An intriguing relation between the length of a region in coding
DNA\index{coding DNA} versus non-coding DNA\index{non-coding DNA}
and a repeatedly higher transport characteristic in coding DNA was
reported in Refs.\ \cite{Shi06,Shi06a}.

The influence of temperature and associated structural fluctuations
of DNA and thus the onsite and hopping parameters has been studied
in the next group of papers. Ref.\ \cite{ConR00} investigates a
polaronic model\index{polaronic model} in which the hopping elements
are influenced by vibrations along the chain. It is shown that a
polaron\index{polaron} can indeed form for reasonable values of
parameters.
Ref.\ \cite{BruGDR00} studies a similar situation, but also takes into
account the rotation between base pairs along the DNA stack. The paper
is actually aimed at charge transfer and proposes that the thermal
fluctuations are the limiting step for site-to-site charge transfer.
Polarons, which have a ``twist'' and can thus model the double-helix
structure\index{double-helix structure} of DNA better, are
investigated in Ref.\ \cite{ZhaGU02}. Non-linear effects are taken
into account and it is shown that these lead to different polaronic
behavior.
In Ref.\ \cite{KomKB02}, it is argued that polaronic transport can
be trapped by the thermal denaturation of poly(G)-poly(C)
DNA\index{poly(G)-poly(C) DNA}. Thermal effects are modelled by an
anharmonic Morse potential\index{Morse potential}.
Semi-empirical quantum-chemical calculations were performed in Ref.\
\cite{YamSHA05} for poly(G)-poly(C) and also for poly(A)-poly(T) DNA
using a polaron\index{polaron} model. Localization lengths of charge
states larger than 2000 base pairs have been computed and it was
shown that significant differences between poly(G)-poly(C) and
poly(A)-poly(T) DNA exist.

Besides temperature, the solution in which DNA is prepared or
measured, its geometry and bend as well as the properties of the
contacts to external leads\index{external leads} will influence the
measured transport characteristics. The influence of disorder for
(\ref{eq:wire-ham}) has been investigated in Ref.\ \cite{ZhaU04}.
Contact effects were studied by Ref.\ \cite{MacTR05} for poly(GACT)
chains. Resonance conditions were identified which show that a
strong coupling to the leads does not always result in larger
conductance.

The simple wire model (\ref{eq:wire-ham}) has also been used for
studies of charge transfer. Briefly, DNA bridges\index{DNA bridges},
containing only AT base pairs, were investigated in Ref.\
\cite{GroBS00} and decay lengths comparable with single-step
tunneling were found. The presence of Kondo bound states
\cite{EndCSP00} leads to long tunneling lengths above $100$nm.
Similarly, time-dependent random hopping strenghts were studied in
Ref.\ \cite{KatL02} and analyzed in a charge transfer context. Last,
a soliton-based\index{soliton} explanation for charge transfer in
long segments of DNA was given in Ref.\ \cite{Lak00}.

\subsection{Importance of base-pairing: two-channel models}
\label{sec-2channel}

A central simplification of the wire model is the description of a DNA
base-pair as a single site. By doing so, one looses the distinction
between a pair with G (or A) on the 5' end of the DNA and a C (or T) on
the 3' side and one where C sits on the 5' and G on the 3', i.e. GC is
equal to CG. This distinction becomes important when considering hopping
between base-pairs, e.g.\ the hopping from GC to AT is different from CG
to AT because of the different size of the DNA bases and thus the
different overlap between G to A and C to A (and similarly for C to T
and G to T) \cite{RosV04}. Furthermore, the relevant electronic states
of DNA (highest-occupied and lowest-unoccupied molecular orbitals with
and without an additional electron) are localised on one of the bases of
a pair only \cite{WanMLR06}. The reduction of the DNA base-pair
architecture into a single site per pair, as in the wire model
(\ref{eq:wire-ham}), is obviously a highly simplified approach.

\begin{figure}
  \centering
  \includegraphics[width=0.9\textwidth]{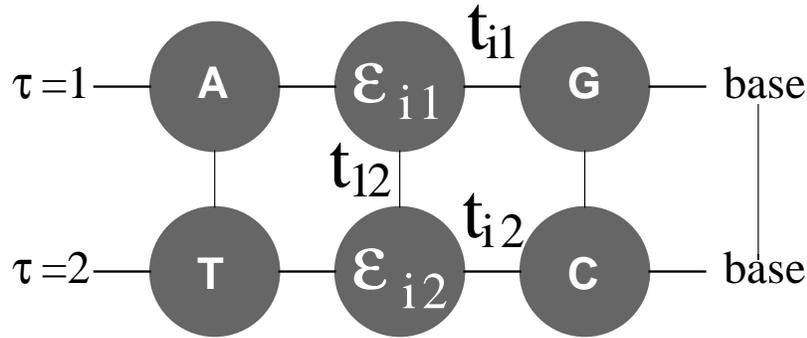}
  \caption{\label{fig-twochannel}
    The two-channel model\index{two-channel model} for electronic transport along DNA. The model
    corresponds to the Hamiltonian (\ref{eq:twochannel-ham}). Electronic
    pathways are shown as lines, whereas the nucleobases are given as
    (grey) circles.}
\end{figure}

This deficiency of the wire model\index{deficiency of the wire
model} may be overcome by modelling each DNA base as an independent
site. The hydrogen-bonding between base-pair is then described as an
additional hopping perpendicular to the DNA stack as shown in Fig.\
\ref{fig-twochannel}.
There are two central branches, linked with one another, with
interconnected sites where each represents a complete base. This {\em
  two-channel} model\index{two-channel model} is a planar projection of the structure of the DNA
with its double-helix unwound, and still without regard for the
backbone. We note that results for electron transfer also suggest
that the transfer proceeds preferentially down one strand
\cite{KelB99}.  The Hamiltonian now reads
\begin{eqnarray}
H_{L} &=& \sum_{i=1}^{L} \left[
 \sum_{\tau=1,2}
    \left( t_{i,\tau}|i,\tau\rangle \langle i+1,\tau| +
    \varepsilon_{i,\tau} |i,\tau\rangle \langle i,\tau| \right) \right.
\nonumber \\
 & & \mbox{ }
+ t_{1,2}|i,1\rangle \langle i,2| \Big]
 + h.c. \label{eq:two-channel-ham}\label{eq:twochannel-ham}
\end{eqnarray}
where $t_{i,\tau}$ is the hopping amplitude between sites along each
branch $\tau=1$, $2$ and $\varepsilon_{i,\tau}$ is the corresponding
onsite potential energy. The new parameter $t_{12}$ represents the
hopping between the two central branches, i.e., perpendicular to the
direction of conduction. As before, we may now attempt to use ab-initio
methods to compute $t_{12}$ or simply model it relative to the strength
of the parallel hopping $t_{i,\tau}$.
For the ordered system with $t_{i,\tau}=t$ and $\epsilon_{i,\tau}=0$, the
two channel model is just a special case of the 2D rectangular system
with spectrum $-2 t_x \cos\left(\frac{\pi k_x}{L_x+1}\right) - 2 t_y
\cos\left(\frac{\pi k_y}{L_y+1}\right)$, $k_x=1, 2, \ldots, L_x$,
$k_y=1, 2, \ldots, L_y$, . Thus we find
\begin{equation}
  \label{eq:two-channel-spectrum}
  E= -2 t \cos\left(\frac{\pi k}{L+1}\right) \mp t_{1,2}
\end{equation}
where the minus (plus) sign correspondes to even,
$\psi_{n,1}=\psi_{n,2}$, (odd, $\psi_{n,1}=-\psi_{n,2}$) states with
the same (opposite) sign for the wave function on each strand.
For random onsite disorder\index{onsite disorder}, the system is
again localized and the localization lengths are known for different
energies and disorder values \cite{RomS04}.

Iguchi was one of the early authors to suggest that a two-leg ladder
model\index{two-leg ladder model} might be a useful starting point
\cite{Igu97}. A band gap like behavior was found in Ref.\
\cite{Yi03}, which also considered the Coulomb
repulsion\index{Coulomb repulsion} between different bases. It was
further shown that for engineered DNA --- modelled as
frustration\index{frustration} --- the band vanishes. Ref.\
\cite{UngS03} used the two-leg ladder model to study the spatial
extent of electronic states\index{extent of electronic states} in
long DNA chains. They found that the extent varies considerably
depending on the sequence, but remains rather small.
Recently, Caetano and Schulz found very large participation
ratios\index{participation ratio} in the two-leg ladder at finite
system sizes \cite{CaeS05}. They speculated that this might indicate
a transition to effectively delocalized states\index{delocalized
states}. But this claim is not expected to hold for longer chains
\cite{SedD06,CaeS06,DiaSSD07}.
The influence of electronic spin and interactions has been studied
in Ref.\ \cite{ApaC05}. This work concentrates on charge transfer
aspects and shows that interaction opens a gap\index{band gap} in
the electronic states of AT and GC pairs.
Further transport properties of Ch22, as well as
$\lambda$-phage\index{$\lambda$-phage} and the histone
protein\index{histone protein}, are investigated in Ref.\
\cite{Yam04} and compared to artificial DNA. It is notable that
while the model used in \cite{Yam04} is a two-leg ladder, the rungs
of the ladder are now modeling not the $\pi$ transport channels but
rather the charge migration along the sugar-phosphate backbone. This
approach is similar to Ref.\ \cite{Igu01}.
Discrete breather-type solutions\index{breather-type solutions}
caused by environmental effects\index{environmental effects} were
studied in a two-leg ladder already in Ref.\cite{ForBL91}. A Morse
potential\index{Morse potential} was used to represent hydrogen
bonding. The breathers\index{breather} were found to be pinned by
the discrete lattice or trapped in defect regions. A similar model
based on the non-linear Schr\"{o}dinger equation\index{non-linear
Schr\"{o}dinger equation} was studied in Ref.\ \cite{WalZ05}, where
the transport of the solitons\index{soliton} was assumed to
propagate along the sugar-phosphate backbone.

\subsection{Backbone effects: The fishbone model}

\begin{figure}
  \centering
  \includegraphics[width=0.9\textwidth]{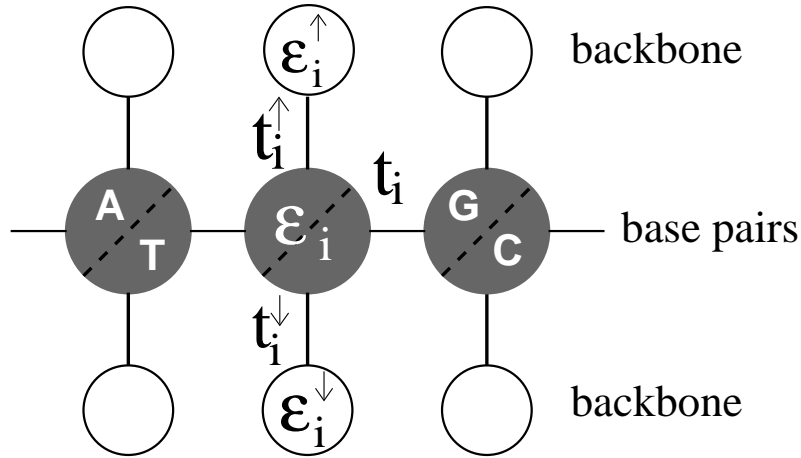}
  \caption{\label{fig-fishbone}
    The fishbone model for electronic transport along DNA corresponding
    to the Hamiltonian given in Eq.\ (\ref{eq:fishbone-ham}). Lines
    denote hopping amplitudes and circles give the central (grey)
    nucleobase pairs and backbone (open) sites.}
\end{figure}

This {\em fishbone model}\index{fishbone model}, shown in Fig.\
\ref{fig-fishbone}, retains the central conduction channel in which
individual sites represent a base-pair. However, these are now
interconnected and further linked to upper and lower sites,
representing the backbone. The backbone sites themselves are
\emph{not} interconnected along the backbone. Every link between
sites implies the presence of a hopping amplitude. The Hamiltonian
for the fishbone model $(H_F)$ is given by
\begin{eqnarray}
H_F &=& \sum_{i=1}^{L}
 \sum_{q=\uparrow,\downarrow} \left(
    -t_{i} |i \rangle \langle i+1|-t_i^q |i,q \rangle \langle i|
    \right.
 \nonumber \\
 & &
\left.+ \varepsilon_i |i \rangle \langle i| + \varepsilon_i^q |i,q
\rangle \langle i,q| \right) + h.c. \label{eq:fishbone-ham}
\end{eqnarray}
where $t_{i}$ is the hopping along the central branch and $t_i^q$ with
$q=\uparrow, \downarrow$ gives the hopping from each site on the central
branch to the upper and lower backbone respectively. We denote the
onsite energy at each site along the central branch by $\varepsilon_i$
and, additionally, the onsite energy at the sites of the upper and lower
backbone is given by $\varepsilon_i^q$, with $q=\uparrow\downarrow$. $L$
is the number of sites/bases in the sequence.
It is easy to see that the existence of the backbone leads to an
effectively renormalized and energy-dependent
disorder\index{renormalized and energy-dependent disorder}
\begin{equation}
  \label{eq:twochannel-disorder}
  \left(\epsilon_n - \frac{{t^\uparrow}^2}{\epsilon_n^\uparrow-E}
    - \frac{{t^\downarrow}^2}{\epsilon_n^\downarrow-E}\right)
\end{equation}
at each base pair $n$ on the $\pi$ stack.
If, we as before for the wire model (\ref{eq:wire-ham}), consider the
ordered situation $t_i=t$, $t^{\uparrow}=t^{\downarrow}$,
$\epsilon_{i}=\epsilon^{\sigma}_i=0$ for $\sigma=\uparrow,
\downarrow$, we find that the energies are now given by
\begin{equation}
  \label{eq:fishbone-spectrum}
  E_{\pm}=-2\cos\left(\frac{\pi k}{L+1}\right) \pm
    \sqrt{t^2 \cos^2\left(\frac{\pi k}{L+1}\right)+2{t^{\uparrow}}^2}
\end{equation}
for $k=1, 2, \ldots, L$.  Hence, there is a highly degenerate state at
$E=0$ corresponding to all the backbone sites and the original
single-band of (\ref{eq:wire-spectrum}) splits into two cosine bands
such that
\begin{equation}
  \label{eq:fishbone-bands}
  E \in \left[ -t -\sqrt{t^2+2 {t^{\uparrow}}^2},
               -t +\sqrt{t^2+2 {t^{\uparrow}}^2} \right]
    \cup
        \left[ t -\sqrt{t^2+2 {t^{\uparrow}}^2},
               t +\sqrt{t^2+2 {t^{\uparrow}}^2} \right].
\end{equation}
In Ref.\ \cite{CunCPD02} it had been shown that this model when applied
to an artificial sequence of repeated GC base pairs, poly(G)-poly(C)
DNA, reproduces experimental data current-voltage measurements when
$t_{i}=0.37$~eV and $t_i^q=0.74$~eV are being used. Therefore, we will
assume $t_i^q = 2 t_{i}$ and set the energy scale by $t_{i}\equiv 1$
for hopping between GC pairs.
Furthermore, since the energetic differences in the adiabatic
electron affinities\index{adiabatic electron affinities} of the
bases are small \cite{WesLPS01}, we choose $\varepsilon_{i}=0$ for
all $i$.

The physics of the fishbone model was first discussed for
poly(G)-poly(C) in Ref.\ \cite{CunCPD02}. In fact, the central sites
of the fishbone are to model the G nucleotide only, with the effect
of the C bases neglected as not so relevant for transport due to
their different onsite HOMO/LUMO energies. The model was then
independently studied by Zhong \cite{Zho03} for random and natural
DNA sequences and he also found an interesting transport enhancing
effect\index{transport enhancing effect} in the band gap\index{band
gap} upon increasing potential disorder. A further study
\cite{KloRT05} revealed that the extent of electronic
states\index{extent of electronic states} in the two bands of the
model can be up to a few dozen base pairs large. Furthermore, upon
adding binary disorder, intended to model adhesion of ions from the
ionic solution in which DNA strands exist, the band gap closes and
the size of initially very well localized band-gap states can be
made to increase substantially \cite{KloRT04}. This effect was also
studied in Refs.\ \cite{GutMC05,GutMC05a} where the system was
coupled to a phonon bath\index{phonon bath}. Here, the band gap was
shown to close with increasing temperature and the temperature
dependence of the charge transmission near the Fermi energy is
exponential.

\subsection{Backbone in a ladder}

\begin{figure}
  \centering
  \includegraphics[width=0.9\textwidth]{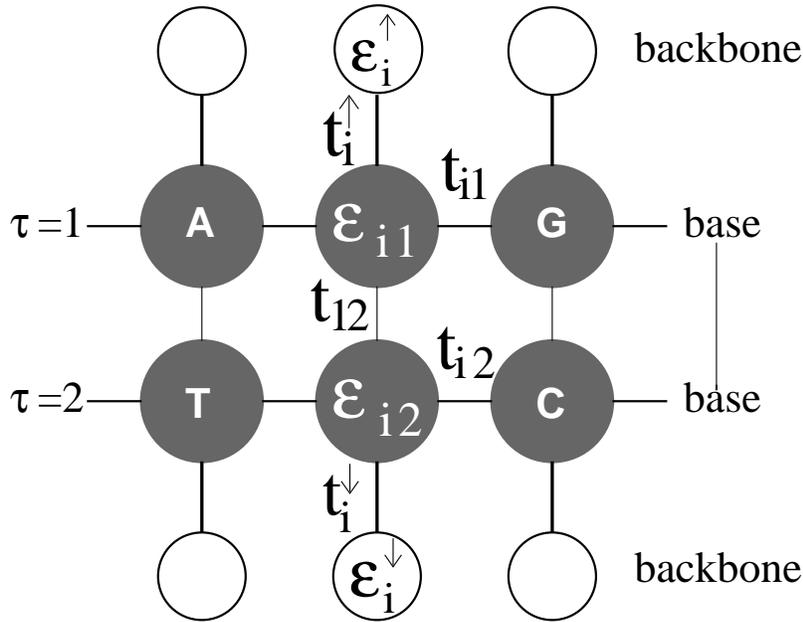}
  \caption{\label{fig-ladder}
    The ladder model for electronic transport along DNA. The model
    corresponds to the Hamiltonian (\ref{eq:twochannel-ham}) and the
    reader should compare the figure to Figs.(\ref{fig-twochannel}) and
    (\ref{fig-fishbone}).}
\end{figure}

Combining the advantages of the fishbone and the two-channel models,
we now model each base as a distinct site where the base pair is
then weakly coupled by the hydrogen bonds. The resulting {\em
ladder} model\index{ladder model} is shown in Fig.\
\ref{fig-ladder}.
There are two central branches\index{two central branches}, linked
with one another, with interconnected sites where each represents a
complete base and which are additionally linked to the upper and
lower backbone sites. The backbone sites as in the fishbone model
are not interconnected. In fact, first principle calculations,
showing that the phosphate molecular orbitals are systematically
below the base related ones, do not favor the possible hopping of
charge carriers between successive phosphate groups along the
backbone.\cite{EndCS04} The Hamiltonian for the ladder model is
given by
\begin{eqnarray}
H_{L} &=& \sum_{i=1}^{L} \left[
 \sum_{\tau=1,2}
    \left( t_{i,\tau}|i,\tau\rangle \langle i+1,\tau| +
    \varepsilon_{i,\tau} |i,\tau\rangle \langle i,\tau| \right) \right.
 \nonumber \\
 & &  \mbox{} + \sum_{q=\uparrow,\downarrow}
    \left( t_i^q |i,\tau\rangle \langle i,q(\tau)|+
    \varepsilon_i^q|i,q\rangle \langle i,q| \right)
\nonumber \\
 & & \mbox{ }
+ t_{1,2}|i,1\rangle \langle i,2|
\Big]
 + h.c. \label{eq:ladder-ham}
\end{eqnarray}
where as before in (\ref{eq:twochannel-ham}) $t_{i,\tau}$ is the
hopping amplitude between sites along each branch $\tau=1$, $2$ and
$\varepsilon_{i,\tau}$ is the corresponding onsite potential energy.
$t_i^q$ and and $\varepsilon_i^q$ as in (\ref{eq:fishbone-ham}) give
hopping amplitudes and onsite energies at the backbone sites. Also,
$q(\tau)=\uparrow, \downarrow$ for $\tau=1, 2$, respectively. The
parameter $t_{12}$ represents the hopping between the two central
branches as for the two channel model (\ref{eq:twochannel-ham}).

For the ordered system with $t_{i,\tau}=t$,
$t^{\uparrow}=t^{\downarrow}$,
$\epsilon_{i,\tau}=\epsilon_{i}^{\sigma}=0$, we find again that the
presence of the backbone sites leads to an effective renormalization
of onsite energies\index{renormalization of onsite energies} along
the two base pair strands with energy-dependent disorder
\begin{equation}
  \label{eq:ladder-disorder}
  \epsilon_{n,\tau}-\frac{{t^{\sigma}}^2}{E-\epsilon^{\sigma}_{\tau}}
\end{equation}
and $(\tau,\sigma)= (1,\uparrow)$ or $(2,\downarrow)$. The
energies for even states are
\begin{equation}
  \label{eq:ladder-spectrum-even}
  E^{+}=\frac{1}{2}\left\{
-t_{1,2}-2t \cos\left(\frac{\pi k^{+}}{L+1}\right)
\pm\sqrt{\left[t_{1,2}+2t \cos\left(\frac{\pi
k^{+}}{L+1}\right)\right]^2+4 {t^{\uparrow}}^2} \right\}
\end{equation}
with $k^{+}=1, 2, \ldots, L$. Similarly, the odd states have energies
\begin{equation}
  \label{eq:ladder-spectrum-odd}
  E^{-}=\frac{1}{2}\left\{
t_{1,2}+2t \cos\left(\frac{\pi k^{-}}{L+1}\right)
\pm\sqrt{\left[t_{1,2}-2t \cos\left(\frac{\pi
k^{-}}{L+1}\right)\right]^2+4 {t^{\uparrow}}^2} \right\}
\end{equation}
and $k^{-}=1, 2, \ldots, L$. Thus we again have two energy bands, with a
slightly smaller gap, given as
\begin{eqnarray}
  \label{eq:ladder-bands}
  E &\in &\left[ -\left(t + \frac{t_{1,2}}{2}\right) -\sqrt{\left(t+\frac{t_{1,2}}{2}\right)^2+ {t^{\uparrow}}^2},
                \left(t + \frac{t_{1,2}}{2}\right) -\sqrt{\left(t+\frac{t_{1,2}}{2}\right)^2+ {t^{\uparrow}}^2}, \right]
    \nonumber \\ &\cup&
        \left[ -\left(t + \frac{t_{1,2}}{2}\right) +\sqrt{\left(t+\frac{t_{1,2}}{2}\right)^2+ {t^{\uparrow}}^2},
                \left(t + \frac{t_{1,2}}{2}\right) +\sqrt{\left(t+\frac{t_{1,2}}{2}\right)^2+ {t^{\uparrow}}^2},  \right].
\end{eqnarray}
In Ref.\ \cite{KloRT05}, electronic transport in this model was
measured by the {\em localisation length} $\xi$, which roughly
speaking parametrises whether an electron is confined to a certain
region $\xi$ of the DNA (insulating behaviour) or can proceed across
the full length $L$ ($\leq \xi$) of the DNA molecule (metallic
behaviour). Various types of disorder, including random potentials,
were employed to account for different real environments.
Calculations were performed on poly(dG)-poly(dC),
telomeric-DNA\index{telomeric-DNA}, random-ATGC DNA and
$\lambda$-DNA\index{$\lambda$-phage}. The authors find that random
and $\lambda$-DNA have localisation lengths\index{localisation
length} allowing for electron motion among a few dozen base pairs
only. A enhancement of localisation lengths\index{localisation
length} similar to the fishbone model (\ref{eq:fishbone-ham}) was
observed at particular energies for an increasing binary backbone
disorder. In Refs.\ \cite{RodRT05,WanMLR06}, the model was used to
study differences in different natural and artificial DNA sequences.
Specifically, promoter sequences\index{promoter sequences} and
sequences known to be repetitive from a biological point of view
were investigated to see whether there were statistically relevant
differences. Using the same sequences as Ref.\ \cite{Shi06a}, no
support for larger $\xi$ values in regions of coding DNA was found.


\section{Conclusions}
\label{sec-conclusions}

In this chapter, we have aimed at giving a review of current models
used for a simplified, tight-binding-based analysis of charge
transport in DNA. While the models can be roughly classified
according to their geometrical structure, many of the presently
available results appear somewhat disjoint and are nearly as widely
spread as in the experimental situation. Let us nevertheless attempt
to identify some common themes. The vast majority of studies
presented here agrees that the transport properties upon including
some degree of energetic disorder --- be it strictly random or
according to some suitable, naturally occurring sequence --- tend
towards the insulating side. Nevertheless, the size of the
electronic states for finite DNA strands might be larger and even
exceed the distance between contacts.  In such a situation, the
experimental results might find finite currents. This finding seems
to be largely independent of the set of on site energies and hopping
strengths chosen.
Also, most studies agree that there are differences between natural DNA
sequences and random DNA with the same ATGC content. However, it is not
clear if these differences are due to the special choice of DNA strands
and simply become statistically irrelevant when other DNA sequences are
considered as well. Thus, a clear correlation between charge transport
and a particular DNA sequence or parts thereof is yet to be discovered.
We emphasize, however, that if such a correlation were to be found, we
would find it useful if it persists across most models reviewed here.

\paragraph{Acknowledgments}
It is a pleasure to thank H.\ Burgert, R.\ Di Felice, D.\ Hodgson,
R.\ Gutierrez, D.\ Porath , R.\ A.\ Remer, S.\ Roche, C.\ T.\ Shih,
A.\ Troisi, M.\ S.\ Turner and E.\ B. Starikov for stimulating
discussions and collaborations on topics related to this chapter.
File hosting has been provided for free by CVSDude.org. G.C.\ would
like to acknowledge support by the the Volkswagen Foundation under
grant No.\ I/78 340, the DFG priority program ``SPP 1243'', the
European Union grant DNAnanoDEVICES under contract No.\
IST-029192-2, the German Israeli Foundation grant No.\ 190/2006, and
the Hans Vielberth Foundation. E.M.\ has been supported by the
Universidad Complutense de Madrid through project
PR27/05-14014-BSCH. R.A.R.\ thankfully acknowledges support by the
EPSRC (EP/C007042/1), the Royal Society and the Leverhulme Trust.




\printindex
\end{document}